\documentclass[twocolumn]{aastex63}
\shorttitle{A Rapidly Rotating Giant in NGC 2243}
\shortauthors{Anthony-Twarog, Deliyannis, Twarog}

\begin{document}


\title{WIYN Open Cluster Study LXXXI. Caught in the Act? The Peculiar Red Giant NGC 2243-W2135}


\author{Barbara J. Anthony-Twarog}
\affiliation{Department of Physics and Astronomy, University of Kansas, Lawrence, KS 66045-7582, USA}
\email{bjat@ku.edu}

\author{Constantine P. Deliyannis}
\affiliation{Department of Astronomy, Indiana University, Bloomington, IN 47405-7105, USA}
\email{cdeliyan@indiana.edu}

\author{Bruce A. Twarog}
\affiliation{Department of Physics and Astronomy, University of Kansas, Lawrence, KS 66045-7582, USA}
\email{btwarog@ku.edu}
%}



\begin{abstract}
High-dispersion spectra for giants through turnoff stars in the Li 6708 \AA\  region 
have been obtained and analyzed in the old, metal-deficient open cluster, NGC 2243. 
When combined with high dispersion data from other surveys, the cluster is found to 
contain a uniquely peculiar star at the luminosity level of the red clump. The giant 
is the reddest star at its luminosity, exhibits variability at a minimum 0.1 mag level 
on a timescale of days, is a single-lined, radial-velocity variable, and has $v\sin{i}$ 
between 35 and 40 km-s$^{-1}$. In sharp contrast with the majority of the red giant 
cluster members, the star has a detectable Li abundance, potentially as high or higher than other giants observed to date while at or just below the boundary normally adopted for Li-rich giants. The observed anomalies may be indicators of the underlying process by which the giant 
has achieved its unusual Li abundance, with a recent mass transfer episode being the 
most probable within the currently limited constraints. 

\end{abstract}



\section{Introduction}

The atmospheric Li abundance (A(Li)\footnote{A(Li)=log$N_{Li}$ - log$N_{H}$ + 12.00}) of 
main sequence stars between $T_{\mathrm{eff}} =$ 5800 K and 7500 K supplies insight into 
the internal structure of the majority of stars populating the red giant region of the color-magnitude 
diagram (CMD) observable today. In broad strokes, the basic outline of the atmospheric evolution
is reasonably well understood. At minimum, convection, to varying degrees, depletes the surface Li abundance 
during the pre-main sequence phase. Stars observed today as A dwarfs would have depleted almost 
no Li, while depletion would have been increasingly severe for increasingly lower-mass dwarfs 
\citep{PI97}. In addition, a large body of evidence suggests that angular momentum loss 
during the main sequence drives mixing and Li depletion for G dwarfs (e.g. \citet{RD95}, F dwarfs 
(e.g. \citet{BO19}), including the striking Li Dip, \citep{BT86}, 
and late A dwarfs alike \citep{DE19}(D19). Thus, nearly all Pop I stars enter the subgiant  
phase with their surface Li depleted by 0.2 to 2 dex or more, depending on mass and other factors. 

Subsequent evolution complicates matters further. There is subgiant dilution due to the severe 
deepening of the surface convection zone, additional non-convective mixing for low-mass giants 
past the luminosity bump on the RGB, and possible mixing due to He-ignition at the tip of the RGB.  
(For an extensive discussion of this complex topic, see D19 and references therein.)

For decades, Li-rich giants (A(Li) $>$ 1.5; see, e.g. \citet{GA19} and references therein) have 
been the exceptions that prove the rule, though the specific mechanism that triggers the Li 
enhancement has remained elusive. Primary candidates include the 
Cameron-Fowler \citep{CF71} mechanism operating within asymptotic giant branch stars, 
enhanced deep mixing operating in first-ascent red giants at the giant branch bump or 
at He-ignition, and planetary engulfment (see, e.g. \citet{CA15, SM18, SO20} and references therein). 

To understand atmospheric Li evolution among low mass stars, an extensive spectroscopic program 
has been underway to survey members of a key set of open clusters from the tip of the giant branch 
to as far down the main sequence as the technology allows. New spectroscopy and/or reanalyses of published
data have been discussed for NGC 752, NGC 3680, and IC 4651 \citep{AT09}, NGC 6253 \citep{AT10, CU12}, 
NGC 2506 \citep{AT16, AT18}, Hyades and Praesepe \citep{CU17}, and NGC 6819 \citep{AT14, LB15, DE19}.
Thanks to a sample size of over 300 stars, the analysis of NGC 6819 led to the discovery of WOCS7017, a 
Li-rich red giant, a class that constitutes $\sim$1$\%$ of the red giant population \citep{GA19, DR19}, 
located significantly redward and fainter than the red giant clump \citep{AT13}. Despite its CMD 
location, later spectroscopic and asteroseismic analysis \citep{CA15, HA17} demonstrated that this 
cluster member is a He-burning clump star of anomalously low mass. 

The purpose of this study is to present preliminary analysis of a red giant with a unique combination of anomalous properties that collectively may have some bearing on the mechanism for retaining and/or enhancing Li abundances in stars well beyond the subgiant evolutionary phase. The star, WEBDA 2135 (W2135), is a member of the older ($\sim$3.6 Gyr \citep{AT05, BR06, TW20}, metal-deficient ([Fe/H] $\sim -0.5$ \citep{AT05, JA11, FR13, AT20} cluster, NGC 2243. 
One prior  estimate of [Fe/H] for W2135 exists, from \citet{Fr02}, a medium-dispersion spectroscopic study of several open clusters.
\citet{Fr02} obtain [Fe/H] $= -0.50$ for W2135, essentially identical to a cluster average of $-0.49$ that includes 8 other giant members. The star is located approximately $2.6\arcmin$ from the cluster center, placing it just within the radius containing 50\% of the cluster members ($r_{50} = 2.76\arcmin$m \citep{CA18}.
As discussed below, W2135 is the reddest star at the level of the clump, a variable, and a single-lined, rapidly rotating, spectroscopic binary. In a cluster where the majority of giants have, at best, 
marginally detectable Li, W2135 falls just at or slightly below the defining boundary for Li-rich red giants.

\section{Observational Data} 
\subsection{Spectroscopy}
Spectroscopic data were obtained using the WIYN 3.5-meter telescope\footnote{The WIYN Observatory was 
a joint facility of the University of Wisconsin-Madison, Indiana University, Yale University, and the National 
Optical Astronomy Observatory.} and the Hydra multi-object spectrograph over 2 consecutive nights in January 2015 
and one night in January 2016. Details regarding the processing and reduction of these spectra will be presented 
in a future paper \citep{AT20}, but follow the pattern laid out in previous cluster analyses such as NGC 6819 
\citep{LB15}. All Hydra stars have been processed, stacked and normalized in identical fashion. The final spectra for 
6 member giants in our bright star configuration are the sums of 3 individual 30-minute exposures, producing 
typical S/N of $\sim 140$. Spectra for the fainter stars in our Hydra sample were constructed by 
taking 9 separate exposures of 53 to 90 minutes in length and stacking them after processing to improve 
the S/N for each cumulative spectrum. W2135 was one of 13 member giants included in the faint star configuration. 
Because of its brightness, adequate S/N was achieved on each W2135 exposure, leading to S/N of 315 for the cumulative Hydra spectrum.

These data were supplemented by spectra collected at the Very Large Telescope (VLT) using the FLAMES fiber-feed assembly to the high resolution UVES spectrograph and the GIRAFFE spectrograph, pixel resolution 16.9 m\AA\ and 50 m\AA\ respectively.  We searched for fully pipeline-processed spectra, initially restricted to spectra with S/N $\geq 70$, that included the spectral range around 670 nm.  The GIRAFFE spectra generally encompass a range 640-680 nm while the higher resolution UVES spectra examined extend further to the blue ($\sim 580$ nm). For the VLT spectra examined and described here, typical S/N are 150. 
 
\subsection{Photometry and Astrometry}
The primary photometric data set is that of \citet{AT05} (ATAT), precision $uvbyCa$H$\beta$ CCD photometry where the number of frames for the giant branch stars ranged typically from 10 for $y$ to 24 for $u$. ATAT used the multicolor indices to identify highly probable cluster members and to convert the high precision ($V$, $b-y$)
for likely members to the traditional ($V$, $B-V$) system defined by earlier, smaller CCD samples \citep{BE91, BO90}. The success of the approach is confirmed by Gaia DR2 \citep{GA18}, where astrometric analysis  \citep{CA18} classifies 26 of the 28 ATAT giant members as definite cluster members, including W2135. 

To make optimal use of all the cluster giants, we have transferred the Gaia DR2 ($G$, $B_{p}-R_{p}$) photometry to the ($V$, $B-V$) system  discussed above using 25 member giants in common (W2135 excluded) to define quadratic relations in ($B_{p}-R_{p}$) 
between the two data sets and merged the two samples. 
The scatter in residuals between the two systems is $\pm$0.007 mag and $\pm$0.009 mag in $V$ and ($B-V$), respectively. The transformation allowed the addition of 15 Gaia member giants not included in the ATAT survey, though one star has only a 20$\%$ membership probability.

As further validation of the DR2 photometry for W2135, we examined the quality factor defined by \citet{Ev18}, the {\tt bp\_rp\_excess} factor. This factor summarizes a photometric consistency check among the $B_{p}$, $R_{p}$ and $G$ passbands.  For photometrically well-behaved sources, this factor should be under $1.3 + 0.06*(B_{p}-R_{p})^2$.  For the case of W2135, this factor is 1.286, well below the color-based criterion of 1.399.

\section{Observational Peculiarities}

As noted earlier, W2135 exhibits a number of photometric and spectroscopic peculiarities relative to
typical stars at a similar state of evolution.

a) Fig. 1 shows the CMD for the NGC 2243 member giants; for later reference, open circles identify the probable clump stars based on their CMD location, while the magenta asterisk tags W2135, clearly the reddest star at a luminosity which coincides well with that of the clump. 
W2135's standout position in the $b-y$ CMD of ATAT is similarly striking and is particularly noted in Fig. 11 of that paper. 
Keeping the lesson of WOCS7017 in NGC 6819 in mind, the extreme color doesn't unambiguously determine if the star is a first-ascent giant or a He-burning member of the red clump. Equally important, while W2135 exhibits variability, discussed next, its CMD color in Fig. 1 is based upon only two sources of photometry, the $b-y$ indices of ATAT and the $B_{p}-R_{p}$ Gaia DR2 indices transformed to a common $B-V$ system. Due to the photometric sequencing for the two data sets, $b$ and $y$ frames always paired back-to-back for ATAT and $B_{p}, R_{p}$ obtained simultaneously for Gaia, the redder CMD position compared to non-variable giants at the same $V$ level cannot be attributed to either photometric error or magnitude differences triggered by filter observations
collected at different phases of the light curve.

\begin{figure}
\figurenum{1}
\plotone{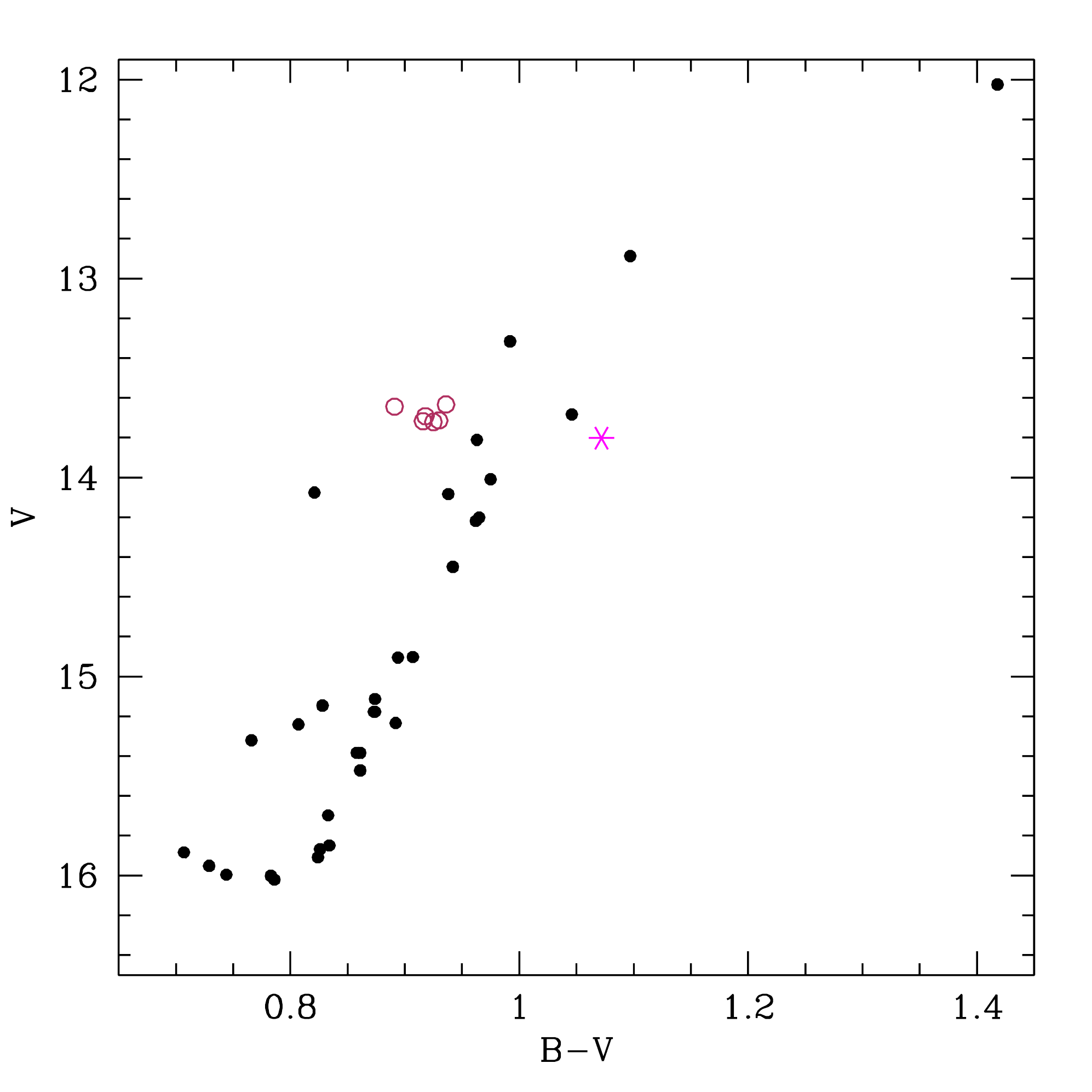}
\caption{Color-magnitude diagram of the giant branch members of NGC 2243. Open circles denote red clump stars while
star W2135 is a magenta asterisk.}
\end{figure}

b) Noting its anomalously redder CMD location and systematically bluer $m_1$ and $hk$ indices, ATAT were the 
first to suggest a potential variable nature for W2135, while expressing concern about possible photometric 
contamination from a faint nearby star. Even though the $uvbyCa$H$\beta$ CCD survey wasn't designed 
with a variable star search 
in mind, the large number of frames in each color obtained in two observing runs 14 months apart clearly 
indicated photometric scatter in all filters at a level 2 to 3 times larger than other stars at the same 
luminosity. The variability range went from 0.09 mag for $y$ to 0.15 mag in $u$, though the larger range 
in $u$ may be a byproduct of a more extensive sample of $u$ measures (24) versus $y$ (9).

The variability was confirmed by \citet{KA06} as part of a search for variable stars with an emphasis on 
finding eclipsing binaries near the cluster turnoff. Of the dozen variables monitored by \citet{KA06}, only
two fell among the cluster giants, V11 (W1496) and V14 (W2135). Because of the luminosity of the giants, 
unsaturated exposures of these stars were achievable during only one run 
%each 
of 5 consecutive nights with a few hours 
of exposures per night. For W2135, a clear night-to-night decline over 4 nights followed by a slight rise in brightness 
on the fifth night, with a full range of at least 0.09 mag, is readily apparent. But, given the limited data, 
all one can say is that if a regular period exists it is more likely to be on a scale of days rather than hours. 
We will revisit the status of W1496 in Section 4.

c) Individual heliocentric stellar radial velocities, V$_{RAD}$, were derived from each summed composite Hydra spectrum 
utilizing the Fourier-transform, cross-correlation facility {\it fxcor} in IRAF\footnote{IRAF is distributed by 
the National Optical Astronomy Observatory, which is operated by the Association of 
Universities for Research in Astronomy, Inc., under cooperative agreement with the National Science Foundation.}. 
In this utility, program stars are compared to stellar templates of similar  $T_{\mathrm{eff}}$ over the full 
wavelength range of our spectra excluding the immediate vicinity of the H$\alpha$ line. Output of the 
{\it fxcor} utility characterizes the cross correlation function, from which estimates of each star's radial 
velocity are easily inferred. 

Continuing the pattern seen among the photometric variability measures, individual spectra of W2135 obtained 
on a given night over a few hours showed no identifiable differences within the uncertainties; the individual 
spectra required negligible wavelength shifts for alignment with each other. However, sets taken on consecutive 
nights of our first run required offsets of $\sim$2.5 km-s$^{-1}$ and stacked spectra taken over runs separated 
by a year required an offset of $\sim$35 km-s$^{-1}$. The simple conclusion is that W2135 is a single-lined 
spectroscopic binary since no evidence for a second set of lines varying in opposition to the bright giant 
is readily apparent, keeping in mind the challenge of detecting the presence of a line set from a potentially much 
fainter companion star against the broad backdrop of the lines in W2135, as discussed below.

d) One of the challenges of working with spectra of W2135, and a potential reason for its absence from  
previous spectroscopic discussions of the cluster, is the exceptional broadness of its lines. Figure 2 illustrates
the exceptional breadth of lines for W2135 by direct comparison between its UVES spectrum and UVES spectra for two giants with similar CMD locations.   
W3618 is positioned in Figure 1 at almost the same $V$ mag level but slightly bluer than W2135. W2648 is
located on the first-ascent red giant branch, fainter and bluer than W2135 and W3618 at ($V, B-V$) = (14.45, 0.94).
All three spectra are approximately normalized to the same level, with two of the spectra vertically offset for greater
visiblity in the plot.  

The spectra illustrate the highest available spectral resolution with a range of S/N, highlighting the potential
difficulties in separating the Li line from the nearby Fe I line at 6707.45 \AA.  The S/N for the three stars' spectra are 103, 192 and 65 for W2135, W3618 and W2648 respectively.
With similar colors, all three stars should exhibit comparable Fe line strengths.  While the two lines are clearly separable in the
highest S/N spectrum of W3618, separation in the noisier spectrum for W2648 is less obvious although this star actually has a strong Li line.  
No separation is possible for the broad lines of W2135 irrespective of S/N or spectral resolution.

\begin{figure}
\figurenum{2}
\plotone{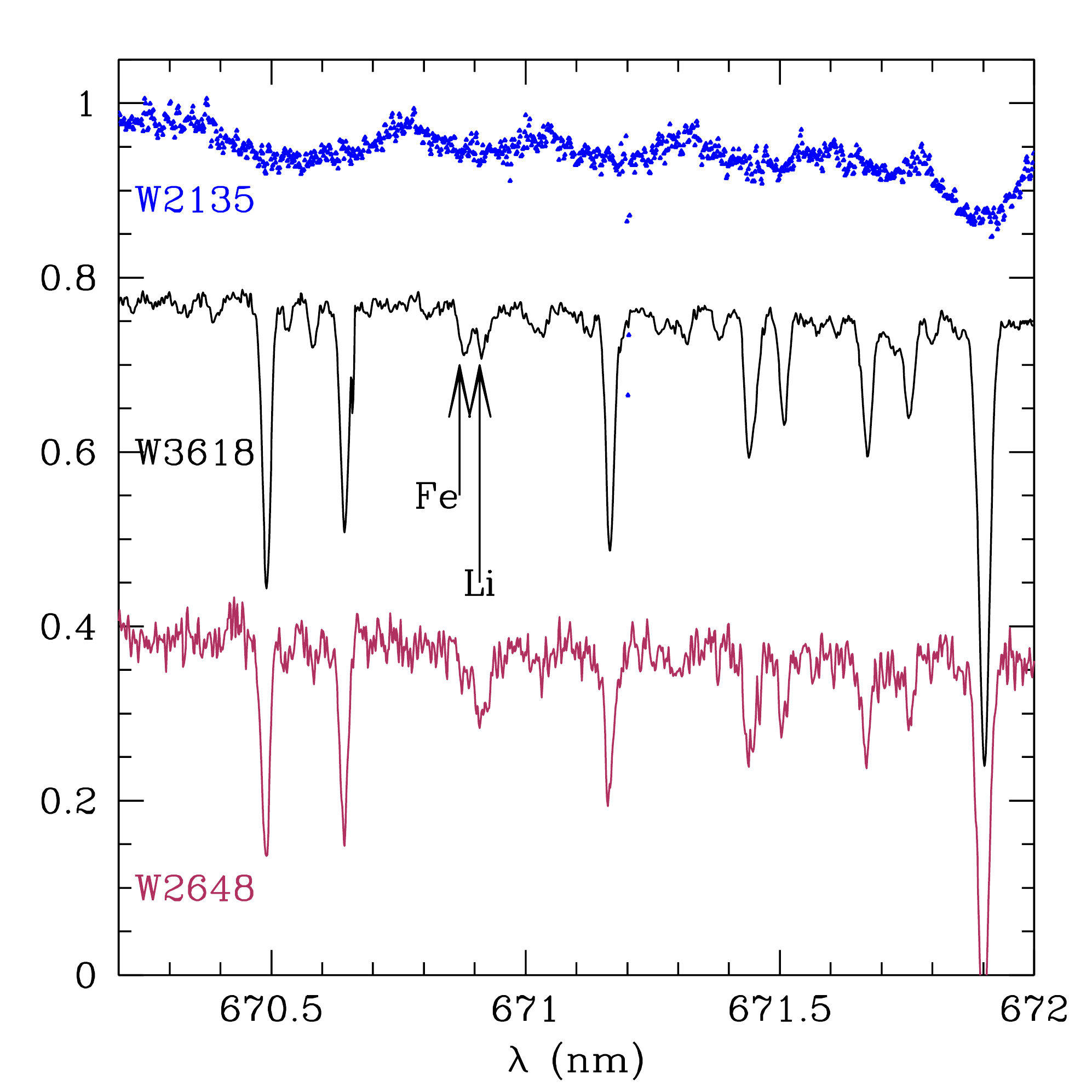}
\caption{UVES spectra of three red giants in NGC 2243 highlighting the region of the Li line. All spectra are 
approximately normalized to a maximum level of 1.0;  the plots for W3618 and W2648 are offset vertically downward 
for visibility.}
\end{figure}

From the procedure for radial-velocity estimation using Hydra spectra discussed above, rotational velocities 
can also be estimated from the cross correlation function full-width (CCF FWHM) using a procedure developed by \citet{ST03}. 
This procedure exploits the relationship between the CCF FWHM, line widths and $v_{ROT}$, using a set of 
numerically ``spun up" standard spectra with comparable spectral types to constrain the relationship.  
For simplicity, $v_{ROT}$ as used here implicitly includes the unknown $\sin{i}$ term. 
From the Hydra spectra for 18 red giant members, W2135 excluded, the mean $v_{ROT}$ is 13.4 $\pm$ 0.9 km-s$^{-1}$ (sem). 
For W2135 the same procedure produces 38.8 km-s$^{-1}$. 

For the Hydra spectra, a majority of the line width for a normal giant is dominated by the instrumental 
line profile since the expectation is that most giants should be spinning in the range of 0 to 5 km-s$^{-1}$. For
simplicity, we will assume that the true $v_{ROT}$ distribution for the giants should average such a small value, 
e.g. close to 1 km-s$^{-1}$, and that the mean observed Hydra value for the red giants is equivalent to the instrumental
profile. Under this assumption, the true $v_{ROT}$ for W2135 is reduced to a minimum of 36.4 km-s$^{-1}$.

The rare nature of these stars is confirmed by the findings of \citet{CA11} for $\sim$1300 field K giants 
where 2.2$\%$ of the stars had $v_{ROT}$ above 10 km-s$^{-1}$ and only 5 stars had $v_{ROT}$ $>$ 30 km-s$^{-1}$. 
\citet{TA15} found only 10 rapidly rotating stars in a sample of 1950 giants, none with $v_{ROT}$ above 25 
km-s$^{-1}$. Three of these were in eclipsing binaries. A percentage comparable to that found by \citet{CA11} 
of 17377 $Kepler$ giants have measurable rotation periods based upon starspot variability \citep{CE17} with 
the intriguing caveat that for low mass, red clump giants, typical of the clump stars in NGC 2243, the 
percentage may rise to 15$\%$. 

With the $v_{ROT}$ derived above and its CMD position (luminosity and $T_{\mathrm{eff}}$), the predicted 
rotation period for W2135 is $\sim$15 days, significantly smaller than found among the typical {\it rotating} 
clump stars in \citet{CE17}. Of 361 red giants with detectable rotation periods, only 17 have periods smaller 
than the W2135 value. It should kept in mind that, unlike periods derived from photometric 
variability, the $\sin{i}$ term within $v{_{ROT}}$ leads to, at best, an upper limit to the rotational period.

e) The primary goals of the NGC 2243 study \citep{TW20, AT20} are to probe the evolution of Li for stars at and above the cluster turnoff
and to place NGC 2243 within the context of clusters of different age and metallicity. 
A comprehensive discussion of the derivation of A(Li) for all NGC 2243  giants, subgiants, and turnoff stars using equivalent width measures and spectrum synthesis will be detailed in \citet{AT20}. For current purposes and clarity, a simple approach using measured and published equivalent widths (EW) can be usefully employed.  
As for previous studies, we employ the SPLOT utility within IRAF to measure equivalent widths and full-widths from Gaussian profile fits.  
  
For previous investigations, we have used a computational scheme developed by \citet{ST03} and employed by \citet{SD04} that interpolates within a model-atmosphere based grid to translate EW values and temperature estimates based on unreddened $B - V$ colors into Li abundances, A(Li).  For spectra inadequate to resolve the contribution of the nearby Fe I line at 6707.45 \AA\ from the Li line at 6707.8 \AA, the computational scheme also numerically subtracts a temperature and metallicity dependent estimate of the Fe line's contribution to the measured EW for the Li line.  With the availability of UVES spectra for some of the giants, it has been possible to improve this color-based estimation of the Fe line contribution, and a correction of ($33.43 * log (B-V)_0 + 16.44$) m\AA\ has been applied to the measured EW for Hydra and Giraffe spectra. EW$\arcmin$ reported for stars with UVES spectra denote the measured EW of the Li line alone.  In this computational scheme, as in future synthesis analyses, the color-temperature prescription of \citet{RA05} is used for giants with an adopted value for E$(B-V) = 0.058$. 

A key feature of the A(Li) abundance estimation scheme crafted by \citet{ST03} is its clear delineation of significant detections, for which the EW for the Li line alone must be at least three times the estimated error in the EW, where the error is estimated from the spectrum resolution, the line full width and the S/N according to a prescription framed by \citet{CA88} and reformulated by \citet{DP93}. As detailed further below and in Fig. 3, only a very few giants have EW$\arcmin$ that will imply a derived A(Li) detection. Moreover, the current computational scheme fails to extend to sufficiently cool temperatures and/or small EW$\arcmin$ values, a failing that will be addressed in \citet{AT20}

Fig. 3 shows the trend of Li EW$\arcmin$, the adjusted EW, with $V$ and ($B-V$). 
Measures from Hydra spectra are shown shown as filled black circles, with open red circles denoting probable red clump stars.
Measurements based upon spectra from GIRAFFE, with resolution four times better than Hydra, are shown as filled blue circles, with open 
blue symbols for red clump stars. Data from the higher resolution UVES spectra are filled teal, with open teal signifying
red clump stars. The one orange filled circle shows the corrected EW$\arcmin$ for W1654, based on the published EW \citep{HI00} and the 
color plotted in Fig. 1.  Stars with both VLT and Hydra spectra are connected by a vertical dashed line. 
Measured EW$\arcmin$ values for W2135 are plotted as a vertical magenta band to illustrate the potential range in EW$\arcmin$ for this star due solely to the difficulty in defining the level of the continuum for these very challenging spectra, whether Hydra or VLT, when attempting to derive the area contained by the line.

We note that the non-Hydra spectral measures include 12 of the 13 stars cooler than 5100 K listed for NGC 2243 in \citet{MA17}. It is assumed that these are
the same 13 stars plotted in Fig. 3 of \citet{CA19} but, unfortunately, no listing of their Li data is provided. Only 12 of the stars are included in our analysis because one of the cool giants listed by \citet{MA17} within NGC 2243, W259, is a Gaia proper-motion and parallax nonmember.

\begin{figure}
\figurenum{3}
\includegraphics[angle=270,width=\columnwidth]{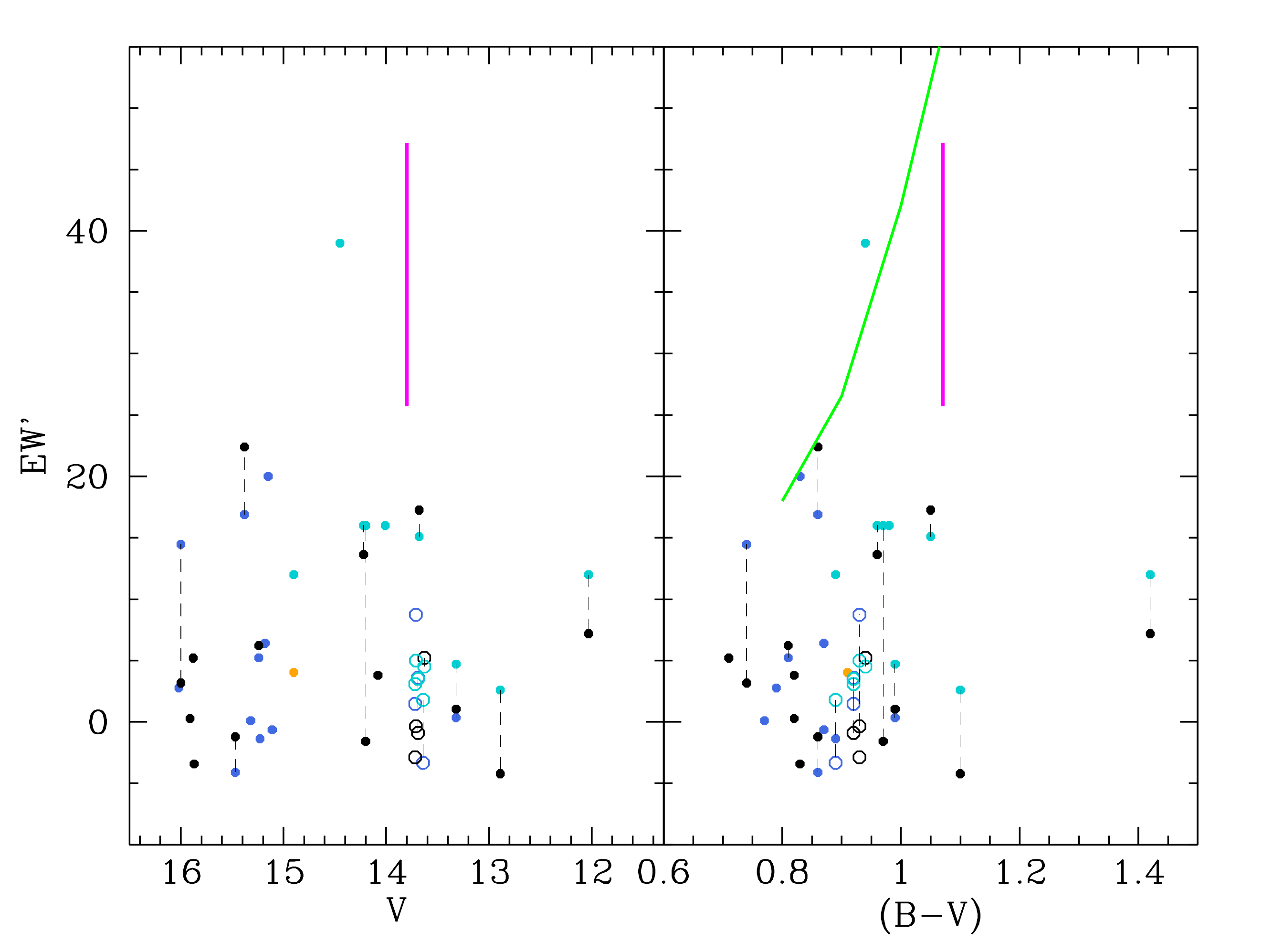}
\caption{Corrected Li EW for all giants from Hydra (black circles), GIRAFFE (blue circles), and UVES (teal circles) 
spectra. Filled orange circle is W1654. Open circles denote red clump stars. Measures for the same star from two sources of
spectra are connected by dashed lines.  The magenta band defines the spread in EW measures for W2135.  The green line shows the expected EW$\arcmin$ values for stars with A(Li) $= 1.0$ as a function of color.}
\end{figure}

Fig. 3 also includes a green band showing presumptive EW$\arcmin$ values as a function of $B-V$ for stars with A(Li) = 1.0. Few of the giants and none of the clump stars approach this abundance value or have EW$\arcmin$ values above 15 m\AA.  The light dashed lines joining measures for the same star illustrate the general agreement among spectroscopic samples with a range indicating uncertainty of $\sim 5$ m\AA. 

One of the few stars with A(Li) potentially above 1.0 is W2648, as noted by \citet{FR13}. Results from the UVES spectrum are shown in Fig. 3 at $(V,B-V,EW$\arcmin$ = 14.45, 0.94, 39).$ Referred to in their survey as K403, \citet{FR13} derive A(Li)$ = 1.5$ for this star. Another star highlighted in their survey, W1261 (K1992), with Fig. 3 coordinates $(15.18, 0.87, 6.4)$, appears less likely to yield a detectable abundance from the Giraffe spectrum, so the abundance A(Li)$ = 1.24$ cited by \citet{FR13} cannot be confirmed. 
Two other stars have EW$\arcmin$ values near the A(Li) $= 1.0$ locus for their color, W2613 (15.15, 0.83, 20) and W611 (15.38, 0.86, 16.9/22.4) with the higher EW$\arcmin$ value for W611 based on the Hydra spectrum. 

Discussion of W2135 is clearly hampered by the large range of measured EW$\arcmin$ values.  This range of measurement uncertainty is not attributable to inadequate S/N but very uncertain continuum placement with respect to the extremely broad lines. We may gain additional insight into the Li abundance for this unique star by comparing its highest S/N spectrum, the lower resolution Hydra spectrum, to a model atmosphere spectrum in Fig. 4. 
We employ \citet{KU92} models for [Fe/H] $= -0.50$, $T_{eff} = 4667$ K, with $log$ $g$ and microturbulent velocities appropriate for its giant branch location. The comparison was completed using the MOOG \citep{SN73} software suite, with spectrum lines broadened to mimic the effect of a large rotational velocity, 40 km/sec.  Three values for A(Li) are shown in the figure, from which a value at or above 1.0 may be safely inferred for this star.

\begin{figure}
\figurenum{4}
\includegraphics[width=\columnwidth]{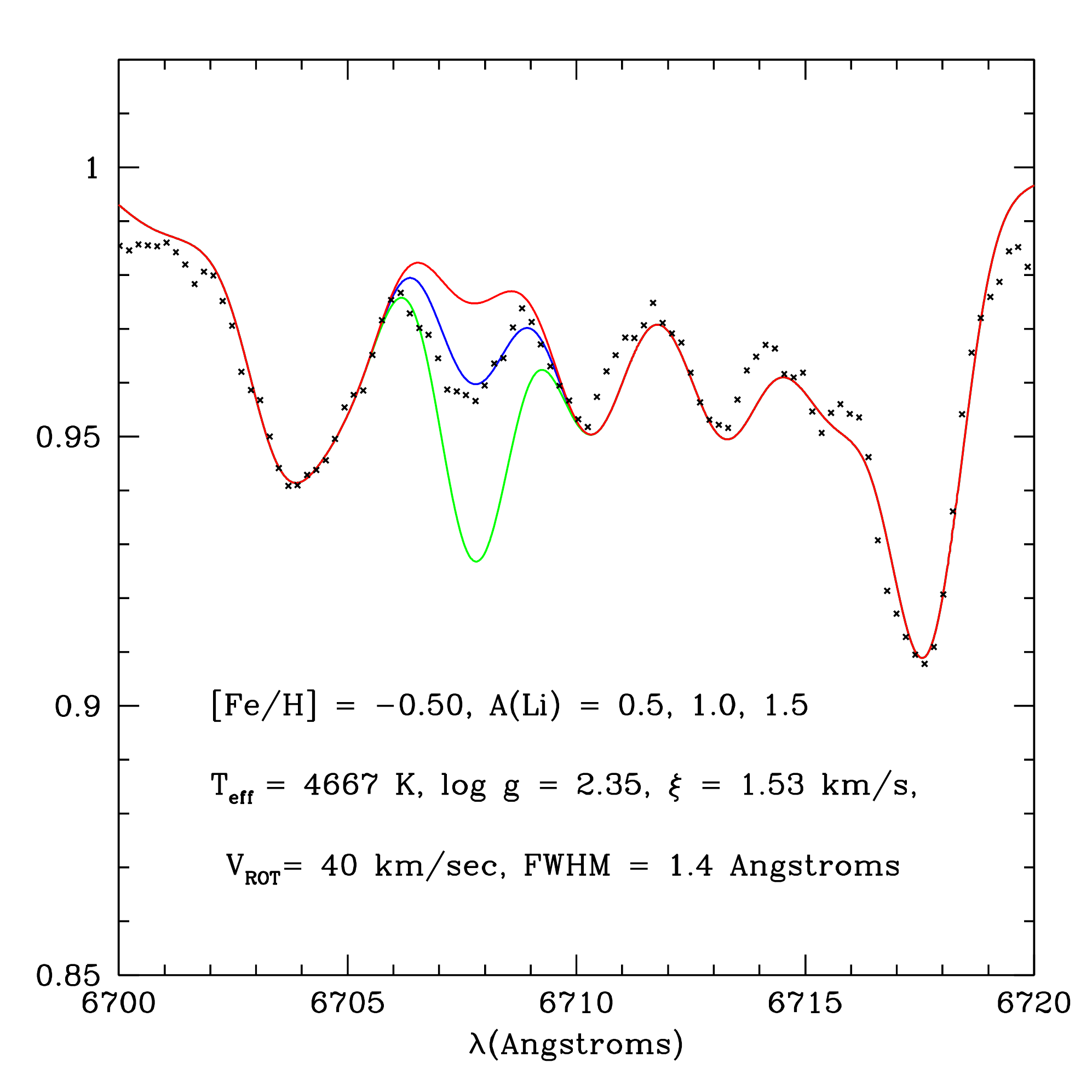}
\caption{Hydra spectrum for W2135 compared to model spectra with A(Li) values of 0.5, 1.0 and 1.5}
\end{figure}

\section{Implications and Future Needs}

%With one exception, the individual anomalies found in W2135 are shared by other stars within the cluster, if notthe red giants.
Taken separately, some of the anomalies - binarity, enhanced lithium, variability - exhibited by W2135 are found in other stars 
in the cluster.  Rapid rotation appears to be unique to W2135.  
Variability of the type seen in W2135 is exhibited by W1496 (V11 in \citet{KA06}). 
Binarity from photometry \citep{KA06} and spectroscopy \citep{KA06b, FR13} has been detected among turnoff stars, but 
has only been inferred from supposedly anomalous CMD positions for some giants. As discussed above, W2135 is close to the 
limit for a Li-rich giant, but its high abundance is matched by one other giant within the cluster. The clear exception 
for W2135 is the rapid rotation rate, coupled with the mix of anomalies not found collectively in any other giant.
It thus appears plausible that some, if not all, of the anomalistic characteristics of W2135 are linked directly or 
indirectly to its exceptional rotation.

Due to lower metallicity and thus a lower mass Li-dip, NGC 2243 bears a strong resemblence to the younger (2.25 Gyr) 
NGC 6819 (D19) in that stars populating the giant branch come from a narrow mass range just outside the Li-dip. Adopting
an age of 3.5 Gyr, [Fe/H] = -0.50 \citep{AT20}, and the same isochrone set \citep{VR06} used to evaluate NGC 6819, 
stars leaving the main sequence in NGC 2243 are just below 1.24 M$_{\sun}$, while the precipitous decline in A(Li) defining
the start of the Li-dip occurs just below 1.21 M$_{\sun}$ \citep{TW20, AT20}. If NGC 2243 follows the
pattern of other clusters, the $v_{ROT}$ distribution of the current turnoff stars originally  
extended to 50 km-s$^{-1}$ or higher. The current mean $v_{ROT}$ among turnoff stars is 
16.2 $\pm$ 1.1 km-s$^{-1}$ (sem), with no star above 22 km-s$^{-1}$, statistically identical to 
NGC 6819 (D19). As noted earlier, the giants, W2135 excluded, have an even smaller mean $v_{ROT}$.

To explain the $v_{ROT}$ of W2135 requires that either it: 

(a) failed to spin down before exiting the main sequence or, 

(b) left the main sequence with low $v_{ROT}$ but was spun up through internal or external evolutionary processes. 

A comprehensive discussion of the many mechanisms proposed to explain rapid rotation (see, e.g. \citet{CA11})
and/or Li enhancement in giants is well beyond the scope of this paper. \citet{CA11} have found that rapid
rotation is more likely to occur among metal-poor and/or lower surface gravity giants, possibly due to the higher 
luminosity and larger radius at a fixed $T_{\mathrm{eff}}$ compared to a metal-rich giant, but triggered by
interaction with a binary companion, possibly in a tidally locked system.  For this reason and given the 
constraints imposed on W2135 by spectroscopy, we will narrow our focus to low mass stars in binary systems. 

If the current $v_{ROT}$ for W2135 is indicative of the initial main sequence $v_{ROT}$, the expected rotational spindown 
and corresponding reduction of A(Li) as the star evolves through the subgiant and giant stages has been severely
curtailed. This might be the case if W2135 is a short-period, tidally locked binary (SPTLB) since 
specific subclasses of these stars are known to retain high rotation and high levels of A(Li) (D19).
Additionally, blue stragglers are commonly assumed to form via mass transfer \citep{MC64} or stellar 
collisions \citep{LE89}, both of which can lead to stellar spinup while still on the main sequence.
The challenge to these solutions is whether or not the evolved star retains the high $v_{ROT}$ as it evolves through
the red giant phase. In the case of blue stragglers, the evidence appears to indicate that stars with these 
anomalies while on the main sequence spin down to a normal, low red giant
$v_{ROT}$ \citep{LE18}, though synchronization triggered in SPTLBs evolving to the giant 
branch cannot be excluded.

If the visible giant left the main sequence with a normal reduction in $v_{ROT}$, its 
current exceptional $v_{ROT}$ might imply a recent spinup via mass transfer while on the giant branch.
Planetary engulfment can spin up a slowly rotating giant to $v_{ROT}$ above 10 km-s$^{-1}$ \citep{CA11}, 
but the exceptional rotation rate of W2135 seems implausible for such a process. Moreover, engulfment 
is more likely to occur at the turnoff and on the lower giant branch while quickly dissipating the effects 
of any Li-enrichment in the upper giant branch \citep{SO20}. When coupled with the significant V$_{RAD}$ 
variation, the limited data point toward a recent mass transfer from a now low-luminosity object of 
stellar mass, possibly a post-AGB/white dwarf star. 

While no other red giant in NGC 2243 studied to date exhibits the same panoply of characteristics as W2135, it should be 
noted that the second variable star tagged by \citet{KA06}, V11 (W1496), has all the non-spectroscopic markings found in 
the primary rapidly rotating candidate. The V11 light curve shows an amplitude of at least 0.05 mag over a range of a few
days, with too few data points to be more specific since the star was only monitored over one 5-night run. In the earlier 
ATAT study, W1496 shows larger scatter than expected for its $V$ magnitude, again in all filters, but wasn't highlighted
as a potential variable since the range in variation wasn't as extreme as that found for W2135. Its location in the CMD, 
if it had been plotted in Fig. 1, would be at ($V, B-V$) = (15.6, 0.73), near the base of the vertical red giant branch, but 
bluer that expected for its $V$ magnitude or, equivalently, a subgiant brighter than expected for its color. It shows the same anomalous 
colors in $m_1$ and $hk$ as W2135, being too blue for its $b-y$, by more than 0.2 mag in $hk$ (see Fig. 8 in ATAT where 
W1496 has ($b-y$, $hk$) = (0.50, 0.49)). Despite the anomalous position, W2135 was retained by ATAT as a member because its 
radial velocity was consistent with membership, a designation later confirmed by both proper motions and parallax from 
Gaia DR2. By contrast, no spectroscopy has
been obtained for W1496 and, surprisingly, the star has not been included to date within the Gaia astrometric data set, which
may explain its absence from more recent spectroscopic surveys. If its membership can be confirmed either spectroscopically 
or astrometrically, W1496 would join the three stars between $V$ = 15 and 16 which clearly lie blueward of the first-ascent 
giant branch, indicative of some form of anomalous origin and/or evolution relative to a normal single star within the
cluster.

We close by noting some potential observational avenues for clarifying the state of the W2135 system. 
Abundance anomalies normally supply signatures for mass transfer events and key ionized lines might provide hints 
about the surface gravity and thus the stellar mass, but the exceptional broadness of the lines in W2135 (Fig. 2) 
illustrates the challenge to this approach. The key to resolving the unusual nature of W2135 lies with the companion. 
A complete light curve would distinguish between eclipses and/or rotationally-defined starspot cycles as the predominant
origin of the variability. If rotation causes variability, the period would constrain the true $v_{ROT}$ and set 
the inclination of the system. If eclipses 
can be detected, one can constrain the properties of the companion and, in conjunction with V$_{RAD}$ 
measures over the orbital cycle, define the mass. 
\newpage
\acknowledgments
We gratefully acknowledge the helpful comments of the referee which led to valuable clarification of the technical aspects of the spectra and their analysis.
NSF support for this project was provided to BJAT and BAT through NSF grant AST-1211621, and to CPD through NSF grants 
AST-1211699 and AST-1909456. Extensive use was made of the WEBDA database maintained by E. Paunzen at the University of Vienna, Austria 
(http://www.univie.ac.at/webda). This research has made use of the services of the ESO Science Archive Facility 
and is based on observations collected at the European Southern Observatory under ESO programme 188.B-3002(V).

\facility{WIYN:3.5m}
\software{IRAF \citet{TODY}, MOOG \citet{SN73}}

\clearpage

\begin{thebibliography}{}
\bibitem[Anthony-Twarog, Atwell, \& Twarog (2005)]{AT05} Anthony-Twarog, B.~J., Atwell, J., \& Twarog, B.~A. 2005, \aj, 129, 872 (ATAT)
\bibitem[Anthony-Twarog et al. (2013)]{AT13} Anthony-Twarog, B.~J., Deliyannis, C.~P., Rich, E., \& Twarog, B.~A. 2013, \apj, 767, L19
\bibitem[Anthony-Twarog et al. (2014)]{AT14} Anthony-Twarog, B.~J., Deliyannis, C.~P., \& Twarog, B.~A. 2016, \aj, 148, 51
\bibitem[Anthony-Twarog et al. (2016)]{AT16} Anthony-Twarog, B.~J., Deliyannis, C.~P., \& Twarog, B.~A. 2016, \aj, 152, 192
\bibitem[Anthony-Twarog et al. (2009)]{AT09} Anthony-Twarog, B. J., Deliyannis, C.~P., Twarog, B.~A., Croxall, K.~V., \& Cummings, J. 2009, \aj, 138, 1171
\bibitem[Anthony-Twarog et al. (2010)]{AT10} Anthony-Twarog, B.~J., Deliyannis, C.~P., Twarog, B.~A., Cummings, J.~D., \& Maderak, R.~M. 2010, \aj, 139, 2034
\bibitem[Anthony-Twarog et al. (2018)]{AT18} Anthony-Twarog, B.~J., Lee-Brown, D.~B., Deliyannis, C.~P., \& Twarog, B.~A. 2018, \aj, 155, 138 
\bibitem[Anthony-Twarog et al. (2020)]{AT20} Anthony-Twarog, B.~J., et al. 2020, in preparation
\bibitem[Bergbusch, VandenBerg, \& Infante (1991)]{BE91} Bergbusch, P.~A., VandenBerg, D.~A., \& Infante, L. 1991, \aj, 101, 2102
\bibitem[Boesgaard, Lum, \& Deliyannis (2020)]{BO19} Boesgaard, A.~M., Lum, M.~G., \& Deliyannis, C.~P. 2020, \apj, 888, 28
\bibitem[Boesgaard \& Tripicco (1986)]{BT86} Boesgaard, A.~M., \& Tripicco, M.~J. 1986, \apjl, 302, 49
\bibitem[Bonifazi et al. (1990)]{BO90} Bonifazi, F., Fusi Pecci, F., Romeo, G., \& Tosi, M. 1990, \mnras, 245, 15
\bibitem[Bragaglia \& Tosi (2006)]{BR06} Bragaglia, A., \& Tosi, M. 2006, \aj, 131, 1544
\bibitem[Cameron \& Fowler (1971)]{CF71} Cameron, A.~G.~W., \& Fowler, W.~A. 1971, \apj, 164, 111
\bibitem[Cantat-Gaudin et al. (2018)]{CA18} Cantat-Gaudin, T., Jordi, C., Vallenari, A., et al. 2018, \aap, 618, 93
\bibitem[Carlberg (2014)]{CA14} Carlberg, J.~K. 2014, \aj, 147, 138
\bibitem[Carlberg, Cunha, \& Smith (2016)]{CU16} Carlberg, J.~K., Cunha, K., \& Smith, V.~V. 2016, \apj, 827, 129
\bibitem[Carlberg et al. (2011)]{CA11} Carlberg, J.~K., Majewski, S.~R., Patterson, R.~J., et al. 2011, \apj, 732, 39
\bibitem[Carlberg et al. (2015)]{CA15} Carlberg, J.~K., Smith, V.~V., Cunha, K., et al. 2015, \apj, 802, 7 
\bibitem[Casali et al. (2019)]{CA19} Casali, G., Magrini, L., Tognelli, E., et al. 2019, \aap, 629, A62
\bibitem[Cayrel de Strobel (1988)]{CA88} Cayrel de Strobel, G. 1988,in {\it IAU Symposium 132, The Impact of Very High S/N Spectroscopy on Stellar Physics}, ed. G. Cayrel de Strobel \& M. Spite (Dordrecht: Kluwer), 345
\bibitem[Ceillier et al. (2017)]{CE17} Ceillier, T., Tayar, J., Mathur, S., et al. 2017, \aap, 505, A111
\bibitem[Cummings et al. (2012)]{CU12} Cummings, J.~D., Deliyannis, C.~P., Anthony-Twarog, B.~J., Twarog, B.~A., \& Maderak, R.~M. 2012, \aj, 144, 137
\bibitem[Cummings et al. (2017)]{CU17} Cummings, J.~D., Deliyannis, C.~P., Maderak, R.~M., \& Steinhauer, A. 2017, \aj, 153, 128
\bibitem[Deepak \& Reddy (2019)]{DR19} Deepak, \& Reddy, B.~E. 2019, \mnras, 484, 2000
\bibitem[Deliyannis et al. (2019)]{DE19} Deliyannis, C.~P., Anthony-Twarog, B.~J., Lee-Brown, D.~B., Twarog, B.~A. 2019, \aj, 158, 163 (D19)
\bibitem[Deliyannis, Pinsonneault, \& Duncan (1993)]{DP93}  Deliyannis, C.~P., Pinsonneault, M.~H., \& Duncan, D.~K. 1993, \apj, 414, 740
\bibitem[Evans et al. (2018)]{Ev18} Evans, D. W., Riello, M., DeAngeli, F. et al. 2018, \aap, 616, A4
\bibitem[Fran\c{c}ois et al. (2013)]{FR13} Fran\c{c}ois, P., Pasquini, L., Biazzo, K., Bonifacio, P., \& Palsa, R. 2013, \aap, 552, A136
\bibitem[Friel et al. (2002)]{Fr02} Friel, E. D., Janes, K. A., Tavarez, M. et al. 2002, \aj, 124, 2693
\bibitem[Gaia Collaboration et al. (2018)]{GA18} Gaia Collaboration, Brown. A.~G.~A., Vallenari, A., et al. 2018, \aap, 616, A1 (DR2)
\bibitem[Gao et al. (2019)]{GA19} Gao, Q., Shi, J.-R., Yan, H.-L., et al. 2019, \apjs, 245, 33
\bibitem[Handberg et al. (2017)]{HA17} Handberg, R., Brogaard, K., Miglio, A., et al. 2017, \mnras, 472, 979
\bibitem[Hill \& Pasquini (2000)]{HI00} Hill, V., \& Pasquini, L. 2000, in I.A.U. Symp. 198, {\it The Light Elements and Their Evolution}, ed. L. Da Silva, J. R. de Medeiros, \& L. Spite (San Francisco, CA: ASP), 293
\bibitem[Jacobson, Friel, \& Pilachowski (2011)]{JA11} Jacobson, H.~R., Friel, E.~D., \& Pilachowski, C.~A. 2011, \aj, 141, 58
\bibitem[Kaluzny et al. (2006a)]{KA06} Kaluzny, J., Krzeminski, W., Thompson, I.~B., \& Stachowski, G. 2006a, \actaa, 56, 51
\bibitem[Kaluzny, Pych, \& Rucinski (2006b)]{KA06b} Kaluzny, J., Pych, W., \& Rucinski, S.~M. 2006b, \actaa, 56, 237
\bibitem[Kurucz (1992)]{KU92} Kurucz, R.~L. 1992, {\it IAU Symp. 149, The Stellar Populations of Galaxies}, ed. B. Barbuy \& A. Renzini (Dordrecht: Kluwer), 225
\bibitem[Lee-Brown et al. (2015)]{LB15} Lee-Brown, D., Anthony-Twarog, B.~J., Deliyannis, C.~P., Rich, E., \& Twarog, B.~A. 2015, \aj, 149, 121
\bibitem[Leiner et al. (2018)]{LE18} Leiner, E., Mathieu, R.~D., Gosnell, N.~M., \& Sills, A. 2018, \apjl, 869, L29
\bibitem[Leonard (1989)]{LE89} Leonard, P.~J.~T. 1989, \aj, 98, 217
\bibitem[Magrini et al. (2017)]{MA17} Magrini, L., Randich, S., Kordopatis, G., et al. 2017, \aap, 603, A2
\bibitem[McCrea (1964)]{MC64} McCrea, W.~H. 1964, \mnras, 128, 147
\bibitem[Pinsonneault (1997)]{PI97} Pinsonneault, M. 1997, \araa, 35, 557
\bibitem[Ram\'{i}rez \& Mel\'{e}ndez (2005)]{RA05} Ram\'{i}rez, I., \& Mel\'{e}ndez, J. 2005, \apj, 626, 465
\bibitem[Ryan \& Deliyannis (1995)]{RD95} Ryan, S.~G., \& Deliyannis, C.~P. 1995, \apj, 453, 819
\bibitem[Smiljanic et al. (2018)]{SM18} Smiljanic, R., Franciosini, E., Bragaglia, A., et al. 2018, \aap, 617, A4
\bibitem[Sneden (1973)]{SN73} Sneden, C. 1973, \apj, 184, 839 
\bibitem[Soares-Furtado et al. (2020)]{SO20} Soares-Furtado, M., Cantiello, M., MacLeod, M., \& Ness, M.~K. 2020, arXiv:2002.05275
\bibitem[Steinhauer (2003)]{ST03} Steinhauer, A. 2003, PhD thesis, Indiana University
\bibitem[Steinhauer \& Deliyannis (2004)]{SD04} Steinhauer, A., \& Deliyannis, C.~P. 2004, \apj, 614, L65
\bibitem[Tayar et al. (2015)]{TA15} Tayar, J., Ceillier, T., Garc\'{i}a-Hern\'{a}ndez, D.~A., et al. 2015, \apj, 807, 82
\bibitem[Tody, D. (1986)]{TODY} Tody, D. 1986, \procspie, 627, 733
\bibitem[Twarog et al. (2020)]{TW20} Twarog, B.~A., Anthony-Twarog, B.~J., Deliyannis, C.~P., \& Steinhauer, A. 2020, {\it Li in the Universe: to Be or not to Be}, \memsai, in press
\bibitem[VandenBerg et al. (2006)]{VR06} VandenBerg, D.~A., Bergbusch, P.~A., \& Dowler, P.~D. 2006, \apjs, 162, 375
 
\end{thebibliography}
\end{document}